\documentclass[aps,pra,reprint]{revtex4-2} 
\usepackage{graphicx}
\usepackage{dcolumn}
\usepackage{bm}
\usepackage{amsmath}
\usepackage{amssymb}
\usepackage{algorithm}
\usepackage{algpseudocode}
\usepackage{xcolor}
\usepackage{tikz}
\usepackage[braket, qm]{qcircuit}
\usepackage{xcolor}
\usepackage{tikz}
\usepackage{qcircuit}
\usepackage{braket}
\usepackage{xcolor}
\usepackage{tikz}
\usepackage{qcircuit}
\usepackage{hyperref}
\usepackage{physics}
\usepackage{float}  
\usepackage{soul}
\begin{document}
\raggedbottom
\preprint{APS/123-QED}
\title{Ancilla-mediated fixed-point quantum search using Grover iterations}

\author{Yash Prabhat}
\author{Snigdha Thakur}
\email{sthakur@iiserb.ac.in}
\affiliation{Department of Physics, Indian Institute of Science Education and Research, Bhopal 462066, India}

\author{Ankur Raina}
\email{ankur@iiserb.ac.in}
\affiliation{Department of Electrical Engineering and Computer Science, Indian Institute of Science Education and Research, Bhopal 462066, India}

\date{\today}

\begin{abstract}
Grover's quantum search algorithm provides a fundamental quadratic speedup for unstructured datasets, reducing query complexity from $\mathcal{O}(N)$ to $\mathcal{O}(\sqrt{N})$.
However, the algorithm's reliance on precise iteration counts leads to the ``soufflé problem,'' where over-rotation results in a sharp decline in success probability. 
This limitation is particularly restrictive when the number of solution states, $M$, is unknown. 
In this work, we present an ancilla-mediated fixed-point quantum search algorithm that achieves robust convergence by mapping the solution amplitude to a dedicated ancilla qubit. 
Unlike existing phase-matching fixed-point methods, our approach utilizes Grover's real-plane reflections, thereby maintaining the intuitive geometric architecture of the original algorithm. 
We demonstrate that this method achieves a success probability of at least $92.6\%$ with a query complexity of approximately $\mathcal{O}(\sqrt{N/M})$, effectively bridging the gap between standard amplitude amplification and robust fixed-point convergence.

\end{abstract}
\maketitle


\section{\label{sec:intro}Introduction\protect\\}
Quantum computing leverages phenomena such as superposition and entanglement to solve problems that remain computationally expensive for classical architectures \cite{NielsenChuang2000, QC_Str_weak, QCsurvey, 10.1098/rspa.1992.0167, doi:10.1137/S0097539796298637, 365700, grover1998}. 
A cornerstone of this field is Grover's search algorithm, which offers a quadratic speedup for searching unstructured datasets \cite{grover1998,  Brassard_2002, Boyer_1998, Brassard1997SearchingAQ}. 
For a dataset of $N$ entries containing $M$ solutions, Grover's algorithm identifies a target state in $\mathcal{O}(\sqrt{N/M})$ iterations, whereas a classical linear search requires $\mathcal{O}(N/M)$ queries \cite{grover1998, Brassard_2002, PhysRevA.64.022307, AmpAmP_H_Peter2000, Brassard1997AnEQ}.

Despite its efficiency, the standard Grover algorithm requires precise control over the number of iterations to avoid the ``soufflé problem'' \cite{Grover_FP}.
Because the success probability follows a sinusoidal curve, over-rotation or under-rotation leads to the selection of non-solution states. 
This sensitivity is particularly problematic in real-world applications where the exact number of solution states, $M$, is unknown. 
To address this lack of robustness, various fixed-point quantum search strategies have been proposed to replace Grover's selective inversions with monotonic convergence mechanisms \cite{Grover_FP, LI2007260, Yoder2014}.


Grover’s initial fixed-point variant utilised $\pi/3$ phase shifts to create attraction basins, ensuring convergence regardless of iteration counts, though it maintained a standard $\mathcal{O}(\sqrt{N})$ scaling \cite{Grover_FP}. 
Subsequent refinements by Yoder \emph{et al.} \cite{Yoder2014} and Li \emph{et al.} \cite{Li_2019} established rigorous convergence bounds via a phase-matching approach; in particular, Li et al. achieve a query complexity $\mathcal{O}(5.643\sqrt{N/M})$ for an unknown number of solutions, giving a quadratic speedup. 

Alternative methodologies for handling unknown $M$ include various trial-and-error techniques \cite{Boyer_1998, YOUNES20081074}. 
The foundational approach proposed by Boyer et al. \cite{Boyer_1998} involves executing Grover's algorithm across multiple trials, with iteration counts chosen randomly from an exponentially increasing interval. 
This stochastic framework ensures that the average success probability remains at least $1/4$ once the algorithm reaches a critical stage, enabling the discovery of a target item with an expected query complexity of at most $4\sqrt{N/M}$. 
While effective, this approach is classified as a randomized application of Grover's algorithm \cite{YOUNES20081074, Younes2004, Li_2019}. 

Closest to the present work are earlier \emph{measurement-based} fixed-point schemes that also employ an ancilla. Grover, Patel, and Tulsi proposed a measurement-driven fixed-point search in which the selective inversion and diffusion operations are controlled by ancilla qubits. 
Irreversible measurement of an ancilla drives the register monotonically toward the target, lowering the error probability as $\epsilon^{2q+1}$ \cite{GroverPatelTulsi2006}; for locating a single marked item. 
However, this construction requires $\mathcal{O}(1/f)=\mathcal{O}(N)$ queries rather than $\mathcal{O}(\sqrt{N})$. 
Mani and Patvardhan subsequently recovered the canonical $\mathcal{O}(\sqrt{N})$ scaling with a fast measurement-based scheme that measures an ancilla and thresholds the ratio of its outcomes as a stopping rule, at the cost of one additional ancilla-controlled query per iteration, an overhead of roughly a factor of two \cite{ManiPatvardhan2011}.
Our construction shares this ancilla-measurement philosophy but differs in three concrete respects: it uses a \emph{single} ancilla purely as a measurable success flag, it retains Grover's real-plane reflections with no phase gates or explicit outcome counting, and it follows a deterministic geometric schedule that both certifies a $\ge 92.6\%$ success floor and provides an explicit termination index for unknown $M$, yielding an expected query complexity $\mathcal{O}(\sqrt{N/M})$.

This work introduces a fixed-point approach that is geometrically distinct from these phase-based and randomized methods.
Instead of phase gates, we utilize Grover’s original reflection operations within an ancilla-mediated architecture \cite{NagyParkZhang2024}. 
This design allows the ancilla qubit to serve as a success flag, which can be measured without destroying the coherence of the data qubits.
This work extends Grover's algorithm \cite{grover1998} to a fixed-point approach using Grover's reflection operations rather than phase gates.
We combine the generality of Grover's approach with a novel fixed-point operator.
The fixed-point operator allows us to measure the ancilla qubit without losing all information about the data qubits.
The measurement of the ancilla qubit indicates whether the solution state is reached.
In conjunction with amplitude-magnification techniques, this fixed-point method allows us to search for an entry in an unstructured dataset with a success probability of $92.6\%$ using an expected query complexity of $\mathcal{O}(\sqrt{N/M})$, where $M$, the number of solution states, is unknown. 

In the worst case, where no solution exists ($M=0$), the algorithm exhausts
   the schedule and terminates after $\mathcal{O}(\sqrt{N})$ oracle calls,
    dominated by the final cumulative count $2^{a-1}\sim\tfrac{\pi}{2}\sqrt{N}$.

The paper is organized as follows. Section \ref{sec:prelim} provides an introduction to the quantum search algorithms. 
Section \ref{sec:algo} provides the probability calculations and an analysis of the query complexity. 
Section \ref{sec:conclusion} compares the existing algorithms with the algorithm in this paper and provides a brief conclusion.


\section{Preliminaries and framework}
\label{sec:prelim}
This work combines the amplitude amplification algorithm \cite{grover1998, Brassard_2002} with a novel approach to the fixed-point algorithm to confidently reach the solution state, thereby successfully completing the search. The said algorithms are given below.

\subsection{Amplitude amplification algorithm}\label{Sec:AmpMag}
The amplitude amplification algorithm is based on the Grover search algorithm, which is a quantum algorithm designed for unstructured search problems \cite{grover1998}. 
The quantum Oracle $\hat {O} $ is used to identify the solution state $\ket {S} $ from the set of all possible dataset states with high probability \cite{grover1998,  Brassard_2002, PhysRevA.64.022307, AmpAmP_H_Peter2000, Brassard1997AnEQ}. 
The Oracle $\hat O$ marks the solution state(s) by flipping their phase. Acting on a computational-basis state $\ket{x}$, it is the unitary
\begin{equation}\label{eq: oracle}
\hat O \ket{x} =
\begin{cases}
-\ket{x}, & x \text{ is a solution},\\
\phantom{-}\ket{x}, & \text{otherwise},
\end{cases}
\qquad
\hat O = \hat I - 2\!\!\sum_{s\in\mathcal S}\ket{s}\!\bra{s},
\end{equation}
where $\mathcal S$ is the set of solutions; equivalently, $\hat O\ket{S}=-\ket{S}$ and $\hat O\ket{R}=\ket{R}$ for the collective solution and non-solution states $\ket{S}$ and $\ket{R}$. 
In this work, the dataset is encoded in $n$ data qubits $q_n$ in initial state $\ket{\psi}$, using Basis/Binary encoding \cite{BinEnc2024,BinaryEncoding} by  a state preparation operator $\hat P$ for some $\theta$ as
\begin{equation}\label{eq: initial}
\hat P\ket{0}^{\otimes n}=\ket{\psi}=\cos\theta\ket{R}+\sin\theta\ket{S}.
\end{equation} 
The state of the ancilla qubit $a$ is shown by $\ket{\psi_a}$ with the initial state taken to be $\ket{0_a}$.
The combined initial state of data qubits and ancilla is represented by $\ket{\psi}\ket{0_a}$. 

\subsubsection*{Amplitude amplification operator}

The amplitude amplification operator ($\hat{G}$) can be defined to amplify the amplitude of the solution state $\ket{S}$.
This is achieved by using Grover's algorithm \cite{grover1998}.
The whole action can be defined as 
\begin{equation}\label{eq: amplitude_ampli_G}
\begin{aligned}
\hat{G}\ket{\psi} &= \hat G(\cos\theta\ket{R}+\sin\theta\ket{S}) \\
&= \cos(3\theta)\ket{R}+\sin(3\theta)\ket{S}.
\end{aligned}
\end{equation}
Here, an operation of $\hat{G}$ has increased the solution state amplitude from $\sin\theta$ to $\sin(\theta+2\theta)$.  
The quantum circuit for the operation can be given by:
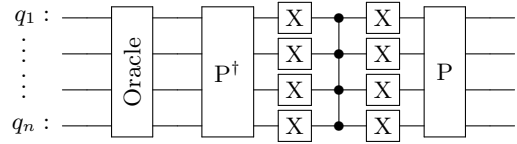
\begin{figure}[H]
    \scalebox{1.0}{
\Qcircuit @C=1.0em @R=0.2em @!R { \\
	 	\nghost{{q}_{1} :  } & \lstick{{q}_{1} :  } & \qw & \multigate{3}{\rotatebox{90}{Oracle}} & \qw & \multigate{3}{\mathrm{P^\dagger}}  & \gate{\mathrm{X}} & \ctrl{1} & \gate{\mathrm{X}} & \multigate{3}{\mathrm{P}} & \qw  & \qw\\
	 	\nghost{  } & \lstick{\rlap{\hspace{-5mm} \raisebox{3mm}{\vdots}}} & \qw & \ghost{\rotatebox{90}{Oracle}} & \qw & \ghost{\mathrm{P^\dagger}}  & \gate{\mathrm{X}} & \ctrl{1} & \gate{\mathrm{X}} & \ghost{\mathrm{P}}  & \qw & \qw\\
	 	\nghost{  } & \lstick{\rlap{\hspace{-5mm} \raisebox{3mm}{\vdots}}} & \qw & \ghost{\rotatebox{90}{Oracle}} & \qw & \ghost{\mathrm{P^\dagger}}  & \gate{\mathrm{X}} & \ctrl{1} & \gate{\mathrm{X}} & \ghost{\mathrm{P}}  & \qw & \qw\\
	 	\nghost{{q}_{3} :  } & \lstick{{q}_{n} :  } & \qw & \ghost{\rotatebox{90}{Oracle}} & \qw & \ghost{\mathrm{P^\dagger}}  & \gate{\mathrm{X}} & \control\qw & \gate{\mathrm{X}} & \ghost{\mathrm{P}}  & \qw & \qw\\
\\ }}

    \caption{Amplitude amplification operator for the solution state $\ket{S}$. Here, Oracle and P represent the quantum circuits of the oracle operator $\hat O$ and the state preparation operator $\hat{P}$, respectively.}
    \label{Fig: amp_amplification_circ}
\end{figure}

There is an increment of angle $2\theta$ by each operation of $\hat G$. After $k$ iterations of amplitude amplification $\hat G ^k$ the state changes to
\begin{equation}\label{eq: Gag}
\hat{G}^k\ket{\psi} =\cos(\theta+2k\theta)\ket{R}+\sin(\theta+2k\theta)\ket{S}.
\end{equation}
Here, the solution state amplitude increased from $\sin\theta$ to $\sin(\theta+2k\theta)$. 
For a known number of solution states $M$, $\theta$ is given by
\begin{equation}\label{eq: theta}
    \theta=\arcsin\left(\sqrt\frac{M}{N}\right).
\end{equation}
 The solution state $\ket{S}$ attains its maximum amplitude when
$\theta+2k\theta=\tfrac{\pi}{2}$. This gives the optimal number of
amplitude-amplification iterations $\epsilon$ as
\begin{equation}
\epsilon=\left\lfloor \frac{\pi}{4}\sqrt{\frac{N}{M}}-\frac{1}{2}\right\rfloor ,
\label{eq: epsilon}
\end{equation}
where $\lfloor\cdot\rfloor$ denotes the floor function and $\theta$ is small.
Next, we look at the fixed-point search method.

\subsection{Fixed point search}\label{sec:FPQSA}

The fixed-point quantum search is a method that fixes a point of convergence, reaching which the search is deemed complete. 
In the proposed algorithm, an ancilla qubit in state $\ket{1}$ acts as a point of convergence for its respective data qubit \cite{Yoder2014, FPQS2017, FPQS2024, Grover_FP}.  For an $n+1$ qubit system, with $n$ data qubits $q_1,q_2,\cdots,q_n$ and an ancilla $a$, we define the state containing $M$ solutions as $\ket{S}$:
\begin{equation}\label{aeq: KetS}
    \ket{S}=\frac{s_1\ket{S_1}}{|S|}+\frac{s_2\ket{S_2}}{|S|}+\cdots+\frac{s_M\ket{S_M}}{|S|},
\end{equation}
where $|S|$ is the normalising factor for the state $\ket{S}$ and $s_i$ is the coefficient of the state $\ket{S_i}$ for all $i\in\{1,2,\cdots,M\}$.
Similarly, we write the rest of the $N-M$ non-solutions as $\ket{R}$:
\begin{equation}\label{aeq: KetR}
    \ket{R}=\frac{r_1\ket{R_1}}{|R|}+\frac{r_2\ket{R_2}}{|R|}+\cdots+\frac{r_{N-M}\ket{R_{N-M}}}{|R|},
\end{equation}
where $|R|$ is the normalising factor for the state $\ket{R}$ and $r_i$ is the coefficient of the state $\ket{R_i}$ for all $i\in\{1,2,\cdots,N-M\}$.

The initial state $\ket{\Psi}$ for the $n$ data qubits and ancilla $a$ for the above basis is given as
\begin{equation}
    \ket{\Psi}= \ket{0_{a}}\left( \cos(\theta)\ket{R} + \sin(\theta)\ket{S} \right)
\end{equation}

A fixed point operator ($\hat{F}(\theta)$) that marks the ancilla qubit $a$ with the probability of the solution state $\ket{S}$ is defined for the orthogonal basis
\begin{equation}\label{aeq: basis}
\begin{aligned}
    \{\ket{0_{a}}\ket{R_1},\ket{0_{a}}\ket{R_2},\cdots,\ket{0_{a}}\ket{R_{N-M}},\\
    \ket{0_{a}}\ket{S_1},\ket{0_{a}}\ket{S_2},\cdots,\ket{0_{a}}\ket{S_M},\\
    \ket{1_{a}}\ket{R_1},\ket{1_{a}}\ket{R_2},\cdots,\ket{1_{a}}\ket{R_{N-M}},\\
    \ket{1_{a}}\ket{S_1},\ket{1_{a}}\ket{S_2},\cdots,\ket{1_{a}}\ket{S_M}\}. 
\end{aligned}
\end{equation}

The column matrix of length $2N$ for the state $\ket{\Psi}$ in this basis would be given as
\begin{equation}\label{aeq: clvector2N}
    \ket{\Psi}=
    \begin{bmatrix}
\cos(\theta) \frac{r_1}{|R|}  \\
\vdots \\
\cos(\theta) \frac{r_{N-M}}{|R|}\\
\\
\sin(\theta)\frac{ s_1}{|S|}  \\
\vdots\\
\sin(\theta) \frac{s_M}{|S|}\\
\\
0\\
\hspace{0.8cm}\vdots \tiny{\text{$N$ times}}\\
 0
\end{bmatrix}.
\end{equation}

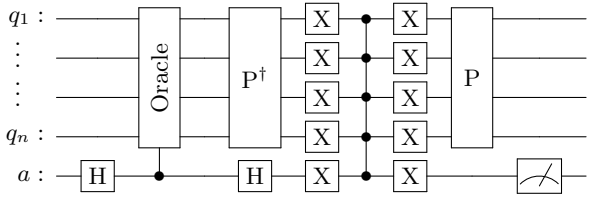
\begin{figure}[h]
    \scalebox{1}{
\Qcircuit @C=1.0em @R=0.2em @!R { \\
	 	\nghost{{q}_{1} :  } & \lstick{{q}_{1} :  } & \qw & \multigate{3}{\rotatebox{90}{Oracle}} & \qw & \multigate{3}{\mathrm{P^\dagger}}  & \gate{\mathrm{X}} & \ctrl{1} & \gate{\mathrm{X}} & \multigate{3}{\mathrm{P}}  & \qw & \qw\\
	 	\nghost{  } & \lstick{\rlap{\hspace{-5mm} \raisebox{3mm}{\vdots}}} & \qw & \ghost{\rotatebox{90}{Oracle}} & \qw & \ghost{\mathrm{P^\dagger}}  & \gate{\mathrm{X}} & \ctrl{1} & \gate{\mathrm{X}} & \ghost{\mathrm{P}}  & \qw & \qw\\
	 	\nghost{  } & \lstick{\rlap{\hspace{-5mm} \raisebox{3mm}{\vdots}}} & \qw & \ghost{\rotatebox{90}{Oracle}} & \qw & \ghost{\mathrm{P^\dagger}}  & \gate{\mathrm{X}} & \ctrl{1} & \gate{\mathrm{X}} & \ghost{\mathrm{P}}  & \qw & \qw\\
	 	\nghost{{q}_{3} :  } & \lstick{{q}_{n} :  } & \qw & \ghost{\rotatebox{90}{Oracle}} & \qw & \ghost{\mathrm{P^\dagger}}  & \gate{\mathrm{X}} & \ctrl{1} & \gate{\mathrm{X}} & \ghost{\mathrm{P}}  & \qw & \qw\\
	 	\nghost{{a} :  } & \lstick{{a} :  } & \gate{\mathrm{H}} & \ctrl{-1} & \qw & \gate{\mathrm{H}}   & \gate{\mathrm{X}} & \control\qw & \gate{\mathrm{X}} & \qw & \meter & \qw  \\
\\ }}
    \caption{The fixed point operator for solution state $\ket{S}$ in $n$ qubits with the measurement instruction on ancilla $a$. Here, Oracle and P represent the quantum circuits of the Oracle operator $\hat {O} $ and the state preparation operator $\hat{P}$.}
    \label{FP_circ}
\end{figure}

The algorithm never leaves the four-dimensional invariant subspace
\begin{equation}
\mathcal{V}=\operatorname{span}\bigl\{\ket{0_a}\ket{R},\,\ket{0_a}\ket{S},\,
\ket{1_a}\ket{R},\,\ket{1_a}\ket{S}\bigr\}.
\end{equation}
On $\mathcal{V}$, in the ordered basis
$\bigl(\ket{0_a}\ket{R},\ket{0_a}\ket{S},\ket{1_a}\ket{R},\ket{1_a}\ket{S}\bigr)$,
the fixed-point operator is the real orthogonal (hence unitary) matrix
\begin{equation}
\hat{F}(\theta)\big|_{\mathcal{V}}=
\begin{pmatrix}
-\cos 2\theta & 0 & 0 & -\sin 2\theta\\[2pt]
-\sin 2\theta & 0 & 0 & \phantom{-}\cos 2\theta\\[2pt]
0 & 0 & 1 & 0\\[2pt]
0 & 1 & 0 & 0
\end{pmatrix},
\qquad \hat{F}^{\dagger}\hat{F}=I_{4}.
\label{eq:F4}
\end{equation}
One verifies directly that
\begin{equation}\label{eq:Fbasis}
\begin{aligned}
\hat{F}(\theta)\,\ket{0_a}\ket{R}
&=-\cos 2\theta\,\ket{0_a}\ket{R}
-\sin 2\theta\,\ket{0_a}\ket{S},\\
\hat{F}(\theta)\,\ket{0_a}\ket{S}&=\ket{1_a}\ket{S},
\end{aligned}
\end{equation}
which fix the action of $\hat F(\theta)$ on the initial state $\ket{\Psi}=\ket{0_a}(\cos\theta\ket{R}+\sin\theta\ket{S})$; the resulting full-state action is stated in Eq.~\eqref{eq: Foptheta} below.
Although $\hat F(\theta)$ depends on $\theta$, its implementation (Fig.~\ref{FP_circ}) uses only the oracle $\hat O$, the state-preparation operator $\hat P$ and its adjoint $\hat P^{\dagger}$, and basic gates, none of which requires the explicit value of $\theta$ or $M$. The circuit of Fig.~\ref{FP_circ} is a product of unitaries and therefore extends $\hat F(\theta)|_{\mathcal V}$ to a unitary on the full $2N$-dimensional space; since the state is confined to $\mathcal V$ throughout, its action off $\mathcal V$ is immaterial.
\paragraph{Gate-level realisation.}
The circuit of Fig.~\ref{FP_circ} implements $\hat F(\theta)$ as the product
\begin{equation}
\hat F(\theta)=\underbrace{\Bigl(\hat P\,\hat R_{0}^{(a=0)}\,\hat P^{\dagger}\Bigr)}_{\displaystyle \hat D}
\;\underbrace{\bigl(H_a\,\text{c-}\hat O\,H_a\bigr)}_{\displaystyle \hat C_S},
\label{eq:Fgates}
\end{equation}
read right-to-left, where $H_a$ is a Hadamard on the ancilla, $\text{c-}\hat O$
is the phase oracle $\hat O$ applied to the data \emph{controlled by the ancilla
being $\ket{1}$}, $\hat P^{\dagger},\hat P$ are (un)preparation, and
$\hat R_{0}^{(a=0)}=I-2\ket{0_a}\!\bra{0_a}\otimes\ket{0}^{\otimes n}\!\bra{0}^{\otimes n}$
places a $-1$ phase on $\ket{0_a}\ket{0}^{\otimes n}$ only, realised by
$X^{\otimes n}$ on the data, a multi-controlled $Z$ with the $n$ data qubits as
$\ket{1}$-controls and the ancilla as a $\ket{0}$-control, and $X^{\otimes n}$.
 
\emph{Step 1 (ancilla flag $\hat C_S$).} Since $H_a\ket{0_a}=\ket{+}$ and
$\text{c-}\hat O$ multiplies the $\ket{1_a}$ branch by $-1$ exactly on solutions,
\[
\hat C_S\ket{0_a}\ket{R}=\ket{0_a}\ket{R},\qquad
\hat C_S\ket{0_a}\ket{S}=\ket{1_a}\ket{S},
\]
because $H_a\ket{+}=\ket{0_a}$ for the untouched $\ket{R}$ branch while
$H_a\ket{-}=\ket{1_a}$ for the sign-flipped $\ket{S}$ branch (phase kickback
converted to a bit flip). Note this requires the oracle to be \emph{ancilla-controlled};
an uncontrolled oracle would phase both branches equally and $\hat C_S$ would
reduce to a global sign.
 
\emph{Step 2 (conditional reflection $\hat D$).} Conjugating the reflection about
$\ket{0}^{\otimes n}$ by $\hat P$ gives the reflection about $\ket{\psi}$, gated on
the ancilla:
\[
\hat D=\ket{0_a}\!\bra{0_a}\otimes\bigl(I-2\ket{\psi}\!\bra{\psi}\bigr)
      +\ket{1_a}\!\bra{1_a}\otimes I .
\]
Using $\braket{\psi}{R}=\cos\theta$, the reflection acts as
\begin{align*}
\bigl(I-2\ket{\psi}\!\bra{\psi}\bigr)\ket{R}
&=\ket{R}-2\cos\theta\ket{\psi} \\
&=(1-2\cos^{2}\theta)\ket{R}-2\sin\theta\cos\theta\ket{S}\\
&=-\cos 2\theta\,\ket{R}-\sin 2\theta\,\ket{S}.
\end{align*}
 
\emph{Composition.} On $\ket{0_a}\ket{R}$, $\hat C_S$ acts trivially (ancilla stays
$\ket{0}$) and $\hat D$ applies the reflection above, giving
$-\cos2\theta\ket{0_a}\ket{R}-\sin2\theta\ket{0_a}\ket{S}$. On $\ket{0_a}\ket{S}$,
$\hat C_S$ produces $\ket{1_a}\ket{S}$, on which $\hat D$ acts as the identity
(ancilla is $\ket{1}$), leaving $\ket{1_a}\ket{S}$. Both agree with
Eq.~\eqref{eq:Fbasis}. The circuit is a product of unitaries, so it
extends $\hat F(\theta)|_{\mathcal V}$ to a unitary on the full $2N$-dimensional
space; since the state never leaves $\mathcal V$, its action off $\mathcal V$ is
immaterial. Each application of $\hat F$ thus uses a single (ancilla-controlled)
oracle call.
The quantum circuit for the operation $\hat F(\theta)$ using $n$ qubits and an ancilla is shown in Fig.~\ref{FP_circ}.

The action of $\hat{F}(\theta)$ on the state $\ket{\Psi}$ is evaluated as
\begin{equation}\label{eq: Foptheta}
\begin{aligned}
    \hat{F}(\theta)\ket{\Psi}&=-\cos\theta\ket{0_{a}}\big(\cos2\theta\ket{R}\\
    &+\sin2\theta\ket{S}\big)+ \sin\theta\ket{1_{a}}\ket{S}.
\end{aligned}
\end{equation}
Here $\hat{F}(\theta)$ maps the state of the ancilla $a$ to $\ket{1_{a}}$ simultaneously coupling the solution state $\ket{S}$ for the data qubits $q_1,q_2,\cdots,q_n$.
This preserves the solution state's amplitude $\sin\theta$ from Eq. (\ref{eq: initial}).
Then, we measure the ancilla qubit $a$ to check for convergence.
If the state $\ket{1_{a}}$ is obtained (probability $\sin^2\theta$), we declare convergence to the solution state $\ket{S}$ for the data qubits. 
If  the state $\ket{0_{a}}$ is obtained (probability $\cos^2\theta$), we get the state
\begin{equation}\label{eq: psieas}
\ket{\Psi'}=\ket{0_{a}}\left( \cos(2\theta)\ket{R} + \sin(2\theta)\ket{S} \right)
\end{equation}
where $\ket{\Psi'}$ is the state of the data qubits and ancilla $a$ after measurement of $a$ in state $\ket{0_{a}}$. 
This allows convergence to the state $\ket{S}$ after multiple measurements on ancilla $a$.


After an arbitrary iteration of the search, the ancilla qubit $a$ is measured, and the state of the data qubits $q_1,q_2,\cdots,q_n$ and $a$ is given as 
\begin{equation}
\ket{\Phi}=\ket{0_a}\ket{\phi}=\ket{0_a}\left(\cos\phi\ket{R}+\sin\phi\ket{S}\right)
\end{equation}\\
with an unknown angle $\phi$, the solution amplitude is $\sin\phi$.
$\hat{F}(\theta)$ maps this amplitude to state $\ket{1_a}$ of ancilla $a$ .
The action of $\hat{F}(\theta)$ on $\ket{\Phi}$ is given as
\begin{equation}
\begin{aligned}
    \hat{F}\ket{0_a}\ket{\phi}=-\cos\phi\ket{0_a}\Bigl(\cos2\theta\ket{R}+\sin2\theta\ket{S}\Bigl)\\ + \sin\phi\ket{1_a}\ket{S}
\end{aligned}
\end{equation}
where $\sin\theta$ is the initial amplitude of solution state $\ket S$.
The $\hat{F}(\theta)$ has successfully mapped the solution amplitude $\sin\phi$ of state $\ket{\phi}$ to the state $\ket{1_a}$ of ancilla $a$. The state $\ket{0_a}$ of $a$ is mapped to $(\cos2\theta\ket{0}+\sin2\theta\ket{1})$, which has preserved the information of the initial state $\ket{\psi}$ from Eq. (\ref{eq: initial}) in the form of angle $2\theta$. This allows repeated measurements of $a$ to reach the solution state confidently. Measurement of the ancilla qubit $a$, changes the state $\ket{\phi}$ of the data qubits as follows. 

\subsubsection*{Measurement}

We measure the ancilla qubit $a$ to be in state $\ket{1_a}$ with probability $\sin^2\phi$.
If $\ket{1_a}$ is obtained, then the data qubits $q_1,q_2,\cdots,q_n$ collapse to the solution state $\ket{S}$, and we stop the search.
After each failure in converging to the solution state (ancilla measured in $\ket{0_a}$), we get back the state $\ket{\Phi_{m}}$ after the measurement of ancilla $a$
\begin{equation}
\ket{\Phi_{m}}=\ket{0_{a}}\ket{\phi_{m}}=\ket{0_{a}}\left( \cos2\theta\ket{R} + \sin2\theta\ket{S} \right) .
\end{equation}
The initial solution and non-solution state probabilities are preserved in the form of $2\theta$.
Thus, the solution state $\ket{S}$ can be searched again in the next iteration.
After the failure of $a$ collapsing to the solution state. $\ket{\phi_{m}}$ is always in the state.
\begin{equation}
\ket{\phi_{m}}=\cos2\theta\ket{R} + \sin2\theta\ket{S}.
\end{equation}
The final implementation combines the amplitude amplification and fixed-point methods above into a single search algorithm.

\section{Algorithm and analysis}
\label{sec:algo}
In this section, we shall look at the final implementation of the fixed-point quantum search algorithm, and its success probability and query complexity calculations.
\subsubsection{Fixed point search algorithm}
The number of iterations $\epsilon_0$ from Eq. (\ref{eq: epsilon}) shall be maximum for $M=1$ as
\begin{equation}\label{eq: t_max_iter}
     \epsilon_0 = \left\lfloor \frac{\pi}{4}\sqrt{N}-\frac{1}{2}\right\rfloor.
\end{equation}
This represents the upper bound on the number of Grover iterations required for the search.
After each failed check the state resets to angle $2\theta$
[Eq.~\eqref{eq: psieas}]. We therefore index the checks by $j=0,1,\dots,a-1$ and
apply the $j$-th check at the geometrically doubling angle
\begin{equation}\label{eq: setK}
\varphi_j = 2^{\,j}\theta , \qquad j = 0,1,\dots,a-1 ,
\end{equation}
where $\theta_{\min}=\arcsin\sqrt{1/N}$ is the smallest angle (worst case $M=1$)
and $a$ is the smallest integer for which the doubling schedule reaches beyond
the first amplitude peak even in this worst case, i.e.
\begin{equation}\label{eq: 2^a}
2^{\,a-1}\theta_{\min} \ge \tfrac{\pi}{2}, \qquad\text{i.e.}\qquad a=\bigl\lceil \log_{2}(\pi\sqrt{N})\bigr\rceil .
\end{equation}
Starting from the reset angle $2\theta$, the angle $\varphi_j$ is reached by
\begin{equation}\label{eq: setK2_j}
g_0 = 0,\qquad g_j = 2^{\,j-1}-1 \quad (j\ge 1)
\end{equation}
amplitude-amplification operations $\hat G$ (note $g_1=0$, since the reset already
sits at angle $2\theta$; $g_0=0$, since the first check acts on the freshly
prepared state at angle $\theta$). The ordered schedule of Grover counts is
$\mathcal K=\{g_0,g_1,\dots,g_{a-1}\}=\{0,0,1,3,7,\dots,2^{\,a-2}-1\}$. As in the
Boyer--Brassard--H{\o}yer--Tapp (BBHT) algorithm~\cite{Boyer_1998}, the schedule
doubles the coverage of the search space, but here in a deterministic
fixed-point context; the first check at $j=0$ gives a high chance of convergence
when $M\sim N$.


These $g_j$ amplifications carry the solution-state amplitude from $\sin 2\theta$ to $\sin(2^{\,j}\theta)$.
Correspondingly, the solution-state amplitude set $\mathcal S_A$ is
\begin{equation}\label{eq: Samp}
\mathcal S_A=\bigl\{\sin\theta,\ \sin 2\theta,\ \sin 4\theta,\ \dots,\ \sin(2^{\,a-1}\theta)\bigr\}.
\end{equation}

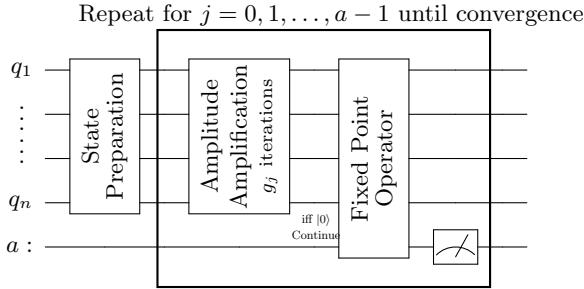
\begin{figure}[h]
    \scalebox{1}{
    \begin{tikzpicture}
        \node at (0,0) {
        \Qcircuit @C=1.0em @R=0.4em @!R { \\
	 	\nghost{q_1} & \lstick{q_1} & \multigate{3}{\rotatebox{90}{\shortstack{State \\ Preparation}}} & \qw & \multigate{3}{\rotatebox{90}{\shortstack{Amplitude \\ Amplification \\ {\scriptsize $g_j$ iterations}}}} & \qw & \multigate{4}{\rotatebox{90}{\shortstack{Fixed Point\\ Operator}}} & \qw & \qw & \qw\\
	 	\nghost{ } & \lstick{\rlap{\hspace{-3mm} \raisebox{1mm}{\vdots}}} & \ghost{\rotatebox{90}{\shortstack{State \\ Preparation}}} & \qw & \ghost{\rotatebox{90}{\shortstack{Amplitude \\ Amplification \\ {\scriptsize $g_j$ iterations}}}} & \qw & \ghost{\rotatebox{90}{\shortstack{Fixed Point\\ Operator}}} & \qw & \qw & \qw\\
	 	\nghost{ } & \lstick{\rlap{\hspace{-3mm} \raisebox{4mm}{\vdots}}} & \ghost{\rotatebox{90}{\shortstack{State \\ Preparation}}} & \qw & \ghost{\rotatebox{90}{\shortstack{Amplitude \\ Amplification \\ {\scriptsize $g_j$ iterations}}}} & \qw & \ghost{\rotatebox{90}{\shortstack{Fixed Point\\ Operator}}} & \qw & \qw & \qw\\
	 	\nghost{q_n  } & \lstick{q_n  } & \ghost{\rotatebox{90}{\shortstack{State \\ Preparation}}} & \qw & \ghost{\rotatebox{90}{\shortstack{Amplitude \\ Amplification \\ {\scriptsize $g_j$ iterations}}}} & \qw & \ghost{\rotatebox{90}{\shortstack{Fixed Point\\ Operator}}} & \qw & \qw & \qw\\
	 	\nghost{{a} :  } & \lstick{{a} :  } & \qw & \qw & \qw & \qw & \ghost{\rotatebox{90}{\shortstack{Fixed Point\\ Operator}}} & \meter & \qw & \qw\\
\\ }};
\draw[thick, black] (-1.2, -1.7) rectangle (3.2, 1.7);
\node[black] at (1.1, 1.9) {Repeat for $j=0,1,\dots,a-1$ until convergence};
\node[black] at (0.9, -0.9) {\scalebox{0.5}{\shortstack{iff $\ket{0}$\\ Continue}}};
\end{tikzpicture}
    }
    \caption{The circuit representing the fixed-point quantum search algorithm.
    At iteration $j$ the state is amplified by $g_j$ Grover operations and then checked by the fixed-point operator; if the ancilla measurement is $\ket{1}$ we have converged to the solution state, otherwise ($\ket{0}$) we continue to the next iteration.
    The convergence probability at the $j$-th measurement is $\sin^2(2^{j}\theta)$, with $j\in\{0,1,\dots,a-1\}$.
    If, after iterating through the schedule, we still have not converged, the solution state does not exist.}
    \label{fig:FPS}
\end{figure}

With set $\mathcal {K} $ and solution amplitude $\mathcal S_{A}$, we define the algorithm \ref{alg: FPS} for the full search operation. 
Fig. (\ref{fig:FPS}) represents the respective quantum circuit. 

\begin{algorithm}[H]
\caption{Ancilla-mediated fixed-point quantum search}\label{alg: FPS}
\begin{algorithmic}[1]
\State Prepare $\ket{\Psi}\gets\ket{0_a}\hat P\ket{0}^{\otimes n}
        =\ket{0_a}(\cos\theta\ket{R}+\sin\theta\ket{S})$
        \Comment{angle $\theta$}
\For{$j=0$ to $a-1$}
    \If{$j\ge 1$}\Comment{amplify from the reset angle $2\theta$ to $2^{j}\theta$}
        \For{$g_j=2^{\,j-1}-1$ iterations}
            \State $\ket{\Psi}\gets\hat G\ket{\Psi}$
        \EndFor
    \EndIf
    \State $\ket{\Psi}\gets\hat F(\theta)\ket{\Psi}$ \Comment{apply the fixed-point check}
    \State Measure ancilla $a$ \Comment{success probability $\sin^2(2^{\,j}\theta)$}
    \If{$a$ is in $\ket{1}$}
        \State Search complete; the data qubits are in the solution state $\ket{S}$.
        \State \textbf{stop}
    \EndIf
    \State \Comment{else the state has reset to $\ket{0_a}(\cos2\theta\ket{R}+\sin2\theta\ket{S})$}
\EndFor
\State \textbf{return} ``no solution exists''
\end{algorithmic}
\end{algorithm}

\subsubsection{Probability calculations and Analysis}
In this subsection, we examine the probability calculations and the worst-case minimum probability of success of the algorithm.
Algorithm~\ref{alg: FPS} gives the probability $p_s$ of converging to the solution state at the $j$-th check as 
\begin{equation}\label{eq14}
p_s=\sin^{2}(2^j\theta)
\end{equation}

where $g_j$ is the number of amplitude-magnification operations $\hat G$ applied
before the $j$-th check [Eq.~\eqref{eq: setK2_j}]. The probability of failing to
collapse to $\ket{S}$ at that check is $\cos^2(2^{\,j}\theta)$.
From Eq. (\ref{eq: theta}, \ref{eq: epsilon} and \ref{eq: 2^a}), a $c,c\le a$ shall exist, satisfying
\begin{equation}\label{eq: jtheta}
\frac{\pi}{4}\le    2^c\theta<\frac{\pi}{2}.
\end{equation}
The probability $p_{c,1}$ of converging within the two consecutive checks at indices $c$ and $c+1$ is
\begin{equation}\label{eq: P_Sj1}
    p_{c,1}=\sin^{2}(2^{c}\theta)+\sin^{2}(2^{c+1}\theta)\,\cos^{2}(2^{c}\theta).
\end{equation}
Let $x=2^{c}\theta$; by Eq.~\eqref{eq: jtheta} the admissible range is $x\in[\tfrac{\pi}{4},\tfrac{\pi}{2})$, and since $2^{c+1}\theta=2x$ this is the single-variable function
\begin{equation}\label{eq: pc1x}
    p_{c,1}(x)=\sin^{2}x+\sin^{2}(2x)\cos^{2}x = 1-\cos^{2}x\,\cos^{2}(2x),
\end{equation}
the second form following from $\sin^{2}\alpha=1-\cos^{2}\alpha$. On $[\tfrac{\pi}{4},\tfrac{\pi}{2})$ the product $\cos^{2}x\,\cos^{2}(2x)$ attains its maximum $0.07407$ at $x=1.15026$, so $p_{c,1}$ attains its minimum $1-0.07407=0.92593$ there, with $p_{c,1}=1$ at the endpoint $x=\tfrac{\pi}{4}$. Because the measurement angles $2^{j}\theta$ double from $\theta$, exactly one of them lies in $[\tfrac{\pi}{4},\tfrac{\pi}{2})$ for every $\theta\in(0,\tfrac{\pi}{2}]$, so a valid index $c$ always exists.

\begin{figure}[h]
    \centering
    \includegraphics[width=1\linewidth]{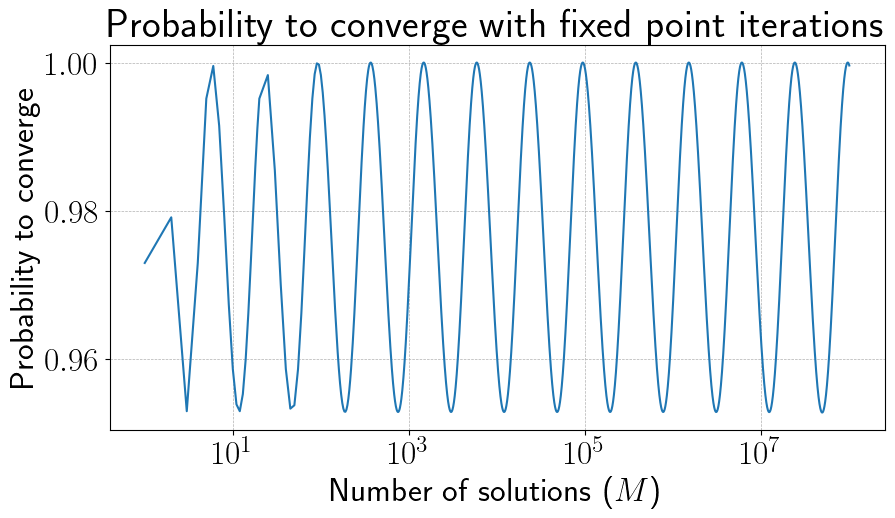}
    \caption{The figure represents the probability of converging to a solution state in $\mathcal{O}(\sqrt{N/M})$ iterations vs $M$ for $N=10^{10}$ by using the fixed-point algorithm. The minimum probability value is numerically 0.95, which is higher than our analytical lower bound of 0.926. }
    \label{fig: Prob VsM}
\end{figure}

The probability of first converging at index $j$ (all earlier checks having failed and reset) is
\begin{equation}\label{eq: PSj}
    P_{S_j}=\sin^{2}(2^{j}\theta)\prod_{m=0}^{j-1}\cos^{2}(2^{m}\theta),
\end{equation}
and the total probability of converging by index $j$ telescopes to
\begin{equation}\label{Eq: Pj}
    P_{j}=\sum_{l=0}^{j}\sin^{2}(2^{l}\theta)\prod_{m=0}^{l-1}\cos^{2}(2^{m}\theta)
         =1-\prod_{m=0}^{j}\cos^{2}(2^{m}\theta),
\end{equation}
where $j$ is the last index in the schedule. Since every factor $\cos^{2}(2^{m}\theta)\le 1$,
\begin{equation}\label{eq: Pjbound}
    P_{j}\ \ge\ 1-\cos^{2}(2^{c}\theta)\,\cos^{2}(2^{c+1}\theta)\ =\ p_{c,1}\ \ge\ 0.9259 ,
\end{equation}
the middle equality being Eq.~\eqref{eq: pc1x} at $x=2^{c}\theta$. Hence $0.9259$ is a rigorous lower bound on the worst-case success probability, independent of $M$.
Figure \ref{fig: Prob VsM} shows these probability values for $\mathcal{O}(\sqrt{N/M})$ iterations in relation to $M$ for $N=10^{10}$.

\subsubsection{Query complexity}
In this subsection, we shall examine the query complexity required to achieve the stated success probability of 0.926.

Each application of $\hat G$ and each application of $\hat F$ is one oracle call,
so the $j$-th check costs
\begin{equation}\label{eq: oc}
o_j = 1+g_j =
\begin{cases}
1, & j=0,\\[2pt]
2^{\,j-1}, & j\ge 1 .
\end{cases}
\end{equation}
The cumulative number of oracle calls to reach the $j$-th check is therefore
\begin{equation}
O_j = \sum_{l=0}^{j} o_l = 2^{\,j}.
\end{equation}

\begin{figure}
    \centering
    \includegraphics[width=1\linewidth]{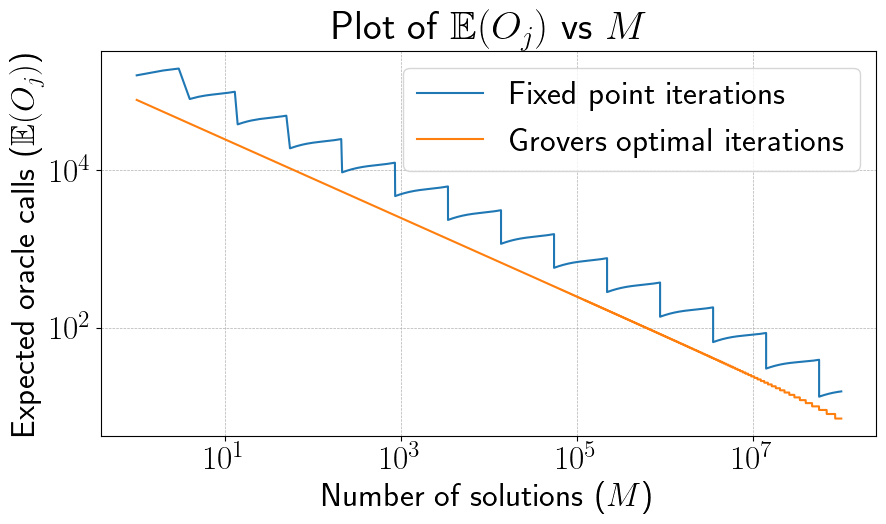}
    \caption{The number of oracle calls required vs $M$ for $N=10^{10}$ by Grover search and the fixed-point algorithm. The mean of the ratio of Grover and FP algorithm iterations is approximately $1.33$.}
    \label{fig: EO VsM}
\end{figure}

By Eq.~\eqref{eq: jtheta}, the two-index guarantee of at least $92.6\%$ is met at the check with index $c+1$; reaching it costs
\begin{equation}\label{eq: OT}
    O_T = O_{c+1} = 2^{\,c+1} < \pi\sqrt{\tfrac{N}{M}},
\end{equation}
where the bound uses $2^{c}\theta<\tfrac{\pi}{2}$ and $\theta\simeq\sqrt{M/N}$. Hence the worst-case complexity of converging to $\ket{S}$ with probability at least $92.6\%$ is $\mathcal{O}(\pi\sqrt{N/M})$.
We have plotted the expected oracle calls $\mathbb{E}(O_j) $ before converging on index $j$ from Eq. (\ref{Eq: Pj}) as
\begin{equation}
    \mathbb{E}(O_j) = \sum_{l=0}^j 2^{\,l}\,\sin^2(2^l\theta)\prod^{l-1}_{m=0} \cos^2(2^m\theta)
\end{equation}
in Fig (\ref{fig: EO VsM}).
$\mathbb{E}(O_j) $ comes out to be approximately $1.33\times \epsilon$ as
\begin{equation}
    \mathbb{E}(O_j) \sim 1.33\,\epsilon,
\end{equation}
for a small $\theta$. 
Further, using the value of $\epsilon$ from Eq. \ref{eq: epsilon}, we get $ \mathbb{E}(O_j)$ as
\begin{equation}
    \mathbb{E}(O_j) \sim \sqrt{\frac{N}{M}}.
\end{equation}

Hence, we show that the algorithm can theoretically reach the success probability higher than 0.926 in less than $\mathcal{O}(\pi\sqrt{N/M})$ oracle calls, thus maintaining the quadratic speedup.
We further numerically demonstrate that the expected number of oracle calls to reach a success probability of 0.95 is approximately $\mathcal{O}(\sqrt{N/M})$ for a dataset of size $N=10^{10}$.

\section{ Discussion and Conclusion }
\label{sec:conclusion}
In this section, we summarize our work, compare it with other existing quantum search algorithms, and provide a brief conclusion.
Unlike standard amplitude amplification, where over-rotation leads to a total loss of the solution state's overlap, our ancilla-mediated approach allows for repeated checkpoint measurements. 
If the ancilla is measured as $\ket{0_a}$, the system remains in a coherent superposition that preserves the initial state information, allowing the search to continue from a known state rather than restarting from scratch.
The algorithm can be modified to serve as an oblivious fixed-point algorithm, marking solution states using calculation operations introduced by Berry \emph{et al.} in 2014 as the oblivious amplitude amplification method \cite{Oblivious_Amp_mag2014}.

Our work has a distinct advantage over existing quantum search methods with an expected query complexity of approximately $\mathcal{O}(\sqrt{N/M})$.
Unlike prior works, our algorithm does not utilize phase gates and is geometrically similar to Grover's algorithm reflection-based approach.
A critical advantage of this reflection-based architecture is its query efficiency: while existing phase-matching fixed-point methods typically require two oracle calls per iteration to achieve convergence, our method uses only one \cite{Yoder2014, Li_2019}. 
This effectively halves the gate depth required for each search step, significantly reducing the impact of decoherence in near-term quantum hardware.
We show a significant complexity advantage over existing work of \cite{Li_2019} (complexity $\mathcal{O}(5.643\sqrt{N/M})$) by achieving a success probability of 92.6\% in $\mathcal{O}(\pi\sqrt{N/M})$ worst case queries. We note that the phase-matching methods of Yoder et al.~\cite{Yoder2014} and Li et al.~\cite{Li_2019} deliver a tunable success probability $1-\delta^{2}$, so their query constants are quoted at a fixed $\delta$; our method instead fixes a $\ge 92.6\%$ guarantee with a simpler, phase-free circuit. 

Our work has a deterministic iteration schedule with a distinct upper bound for search termination, which differentiates it from existing randomized approaches. 
\begin{table*}[ht]
\centering
\renewcommand{\arraystretch}{1.3}
\caption{A comparison of quantum search algorithms}
\label{tab:algorithm_comparison}
\begin{tabular}{|l|l|c|l|l|}
\hline
\textbf{Algorithm} & \textbf{Mechanism} & \textbf{Convergence} & \textbf{Query Complexity} & \textbf{Success Prob.} \\ \hline
Grover F.P. \cite{Grover_FP} & $\pi$/3 Phase matching F.P. & Deterministic & $\mathcal{O}(\frac{\ln(1/\delta)}{\lambda})$ \textsuperscript{b}  & $\frac{1}{4}\lambda_0^2+\frac{1}{16}\lambda_0^3$ \textsuperscript{a}  \\ \hline
Boyer et al. \cite{Boyer_1998, Li_2019} & Trial-and-error  & Randomized & $\mathcal{O}(4\frac{1}{\sqrt{\lambda_0}})$ \textsuperscript{b}   & Arbitrary (at least 1/4) \\ \hline
Yoder et al. \cite{Yoder2014} & Phase matching F.P.& Deterministic  & $\mathcal{O}(\frac{\ln(2/\delta)}{\sqrt\lambda})$  \textsuperscript{b}  & $1-\delta^2$ \textsuperscript{c} \\ \hline
Li et al. \cite{Li_2019} & Randomized-phase matching hybrid  & Deterministic & $\mathcal{O}(5.643{\frac{1}{\sqrt{\lambda_0}}})$ \textsuperscript{a}  &  $1-\delta^2$ \textsuperscript{c} \\ \hline
This work & Ancilla mediated real plane F.P.& Deterministic  & $\mathcal{O}(\pi{\frac{1}{\sqrt{\lambda_0}}})$ \textsuperscript{a,d} & $\ge 92.6\%$ \\ \hline
\end{tabular}
\begin{flushleft}
\textsuperscript{a} $\lambda_0$ represents the fraction $M/N$. (\ref{eq: epsilon}).\\
\footnotesize{\textsuperscript{b} Requires a known lower bound $\lambda$ for the fraction $M/N$ \cite{Yoder2014}}\\
\footnotesize{\textsuperscript{c} $\delta$ is the optimal parameter to minimize the expected query complexity \cite{Yoder2014,Li_2019}}\\
\footnotesize{\textsuperscript{d} Expected number of oracle calls are approximately $\mathcal{O}(\frac{1}{\sqrt{\lambda_0}})$}.
\end{flushleft}
\end{table*}
Table (\ref{tab:algorithm_comparison}) gives the mechanism, convergence, query complexity and success probability of existing search algorithms and ours. 
Further, our work provides clear quantum circuits for easy implementation of the algorithm for future use.

The fixed-point quantum search paradigm achieves robust convergence; by decoupling the success probability from the precise knowledge of $M$, we have circumvented the soufflé problem. 
The algorithm achieves a stable success rate of $\ge 92.6\%$ while ensuring that subsequent iterations do not degrade solution quality, a common failure mode in unstructured search.
The fixed-point search has a query complexity of $\mathcal{O}(\sqrt{N/M})$, similar to Grover's search, without requiring the optimal number of iterations. 
This solves a major problem with the practical applications of Grover's search algorithm for searching an unknown dataset while maintaining its core architecture.
This makes this method geometrically distinct from prior phase-based fixed-point approaches while maintaining similar or better time-complexity performance.
In the worst case, with an unknown number of solution states $M$, the proposed algorithm requires at most $\mathcal{O}(\pi\sqrt{N/M})$ oracle calls.
The expected number of oracle calls to converge to a solution is $\mathcal{O}(\sqrt {N/M})$ with a probability of at least 92.6\%.
This architecture attempts to solve the search problem inspired by Grover's algorithm, achieving $\mathcal{O}(\sqrt{N/M})$ complexity when $M$ is unknown.
Future work could explore the integrating fixed-point operator into Adaptive Search frameworks for quadratic binary optimization, where the number of solutions changes dynamically and is inherently unknown \cite{NagyParkZhang2024}.

\begin{acknowledgments}
The authors thank IISER Bhopal for providing the computational resources to conduct this research.
\end{acknowledgments}

\bibliography{apssamp}

\end{document}